# Computational Analysis of Inspiratory and Expiratory Flow in the Lung Airway

Peshala P. T Gamage[*], Fardin Khalili[1], Azad Md. K[1], Hansen A Mansy[1]

[1]University of Central Florida, Department of Mechanical and Aerospace Engineering, Biomedical Acoustic Research Lab, Orlando, FL 32816, USA

## ABSTRACT

Inspiratory and expiratory flow in a multi-generation pig lung airway was numerically studied at a peak tracheal flow rate corresponding to a Reynolds number of 1150. The model was validated by comparing velocity distributions with previous measurements for a simple airway bifurcation. Simulation results at different cross sections of the airway tree provided detailed maps of the axial and secondary flow patterns. Flow at the main bifurcation and in many other bifurcations showed complex secondary flow structures. The flow morphology in the pig airways differed from that of simplified bifurcation airway models and that of humans, which is likely due to the large differences in the airway geometry of the different species. The inspiratory pressure drop was calculated, and simulation results suggested that the viscous pressure drop values were comparable to earlier studies in human airway geometries. The reported differences between pig and human airways need to be taken into consideration when generalizing results of animal experiments to humans.

## 1. INTRODUCTION

The airflow dynamics in lung airways is vital to the studies of respiratory illnesses like COPD, asthma, and pneumothorax. In addition, understanding the airflow is important in the studies of drug delivery, particle deposition and sound generation in the airways. Earlier studies [1, 2] discussed flows in simplified airway geometry models. While the results of these studies discussed the fundamental flow patterns, velocity distributions in relatively simple geometries, recent studies in computational fluid dynamics(CFD) [3, 4] provided important details of the characteristics of airflow in more complex airway geometries and showed that flow patterns in simplified geometries can significantly differ from actual cases [3]. Earlier studies [5, 6] often used animal models to understand airflow characteristics in airway tree due to difficulties associated with human in-vivo studies. While some studies[7] discussed respiratory mechanics and relevant airflow phenomena in the pig airways, other recent studies discussed the geometrical aspects of domestic pig airways [8] and the differences from that of humans.
The current study performed numerical CFD analysis of airflow dynamics for a realistic pig airway geometry. The methods were fist validated using data from previous experimental studies [1, 2] of simplified airway models. Then the flow behaviour in pig airway is analysed at peak inspiratory and expiratory flow rates and results are discussed in relation to previous studies in human airways.

## 2. MATERIALS AND METHOD

The flow is simulated using Reynolds Averaged Navier stokes (RANS) SST k-ω model in Star CCM+ software [9]. The airway geometry of domestic pig (Yorkshire-Landrace breed, weight~40kg) was obtained from a CT scan of a previous study [10]. The geometry was meshed in Star CCM+ which contained ~ 2 million cells after a grid independence study and Y+ value was maintained less than 1. More details about the mesh can be found in a similar previous study[7].Mass flow rate boundary condition was imposed at the inlet and zero static pressure was imposed at outlets. Non-slip boundary condition was imposed at walls. A turbulent intensity of

*Corresponding Author: peshala@knights.ucf.edu

5% was specified at the inlet to account for the turbulence generated from upper airways[7]. The fluid properties were such that: the density, ρ = 1.2 kg.m-3 and dynamic viscosity, µ= 1.85×10-5 kg.m/s.

## 3. RESULTS AND DISCUSSION

### 3.1 Validation of the Flow in Simplified Airway Geometries

Velocity distribution results were validated by comparing them with LDV (Laser Doppler Velocity) measurements [1] of inspiratory flow in a single bifurcation. Fig. 1 shows the simulated and measured velocity distributions at different sections. Good agreement between the experimental and numerical results was observed.

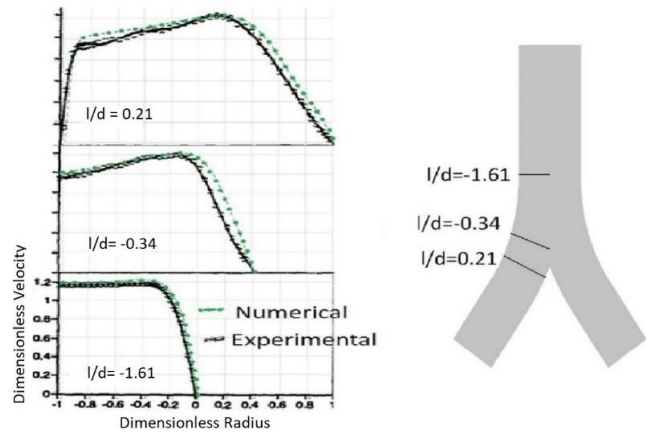

**Fig**. 1 Comparison between simulated[7] and measured [1] velocity distributions at different bifurcation planes for an inlet Re=1500

### 3.2 Flow in Pig Airway

*Axial and secondary velocity distribution*

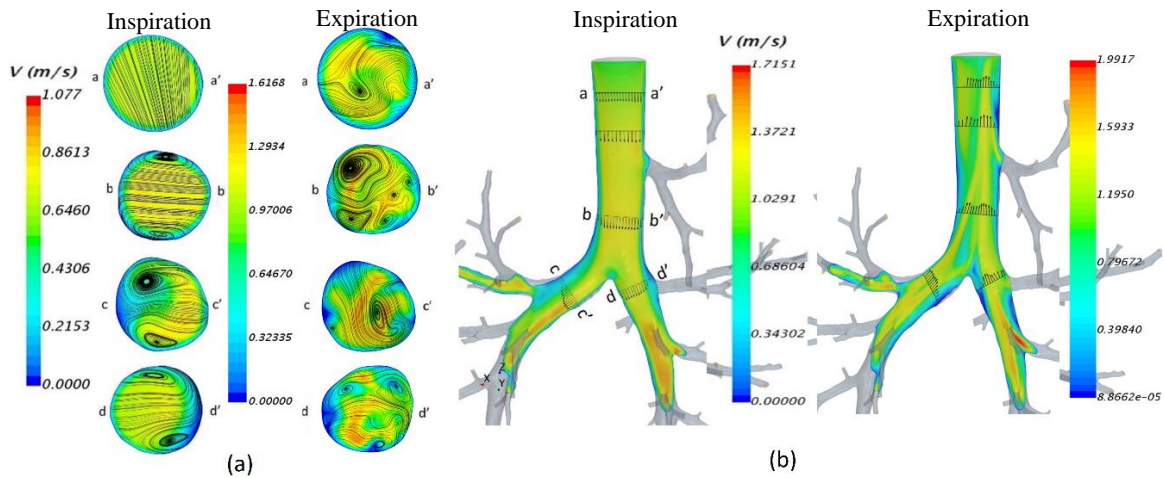

**Fig**. 2 (a) secondary velocity streamlines with axial velocity contours at different sections (b) axial velocity at a cross section of pig airway for inspiratory and expiratory flow

The axial flow velocity and secondary velocity streamlines at different sections for both inspiratory and expiratory flow are shown in Fig. 2. Two counter rotating vortices due to Dean's flow were seen in the cross sections (section c-c' and d-d') at main bronchi for inspiratory flow. Development of these counter rotating vortices was visible at section b-b' before the flow bifurcates at the carina. These vortices were not symmetric compared to the vortices observed in previous studies for simplified human airway bifurcations during inspiration [2]. The irregular geometry of the realistic airways probably caused the vortices to be non-symmetric. The expiratory results showed more complex flow structures compared to inspiratory flow. Several secondary vortices were observed in sections b-b', c-c'and d-d' while one secondary vortex was visible upstream to tracheal bronchus at section a-a'. Secondary vortex formation at section b-b' was significantly different from 4 secondary vorticities observed at trachea during inspiration in human lung airways [2]. These different flow pattern will likely affect particle deposition, turbulence and flow-generated breath sounds.

*Flow rate distribution*

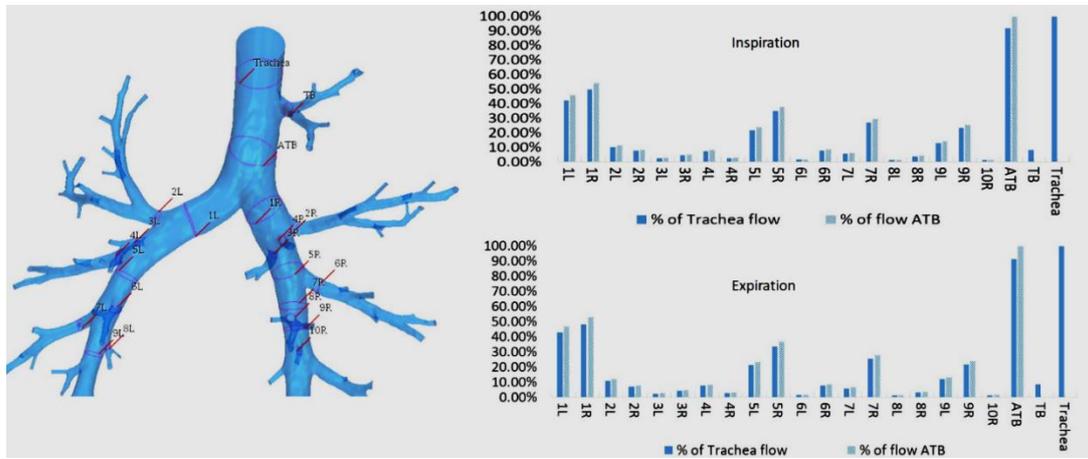

**Fig**. 3 Flow rate distribution got inspiration and expiration

The flow rate distribution results at several sections of the pig airway for inspiratory and expiratory conditions are shown in Fig.3. The ratios are calculated as a fraction of flow entering at trachea and as a fraction of flow after tracheal bronchus (ATB). While distribution results for inspiration and expiration show very similar results, overall distribution flow results indicate that majority of the flow remains in the major daughters (sections: 1L,5L,9L,1R,5R,7R,9R) compared to minor daughters. Higher flow percentage in the major daughters is likely due to monopodial bifurcation morphology in pig airway [8], where major daughters have large diameters and small deflection angles compared to minor daughters. Results also showed that 8% of inlet flow is consumed by the tracheal bronchus (TB) while 49.8% and 42.1% of the flow is consumed by right and left lung respectively. When the flow ratios are calculated as a fraction of flow at ATB, 54% and 46% of flow ratios were consumed by right and left lung. These values are close to the flow ratios consumed by right and left lungs of human calculated in a previous study[11]. This is likely because the bifurcation at the main carina involves branch diameters and angles similar to humans.

*Pressure drop study*

The pressure drops were calculated at each generation for inspiration and compared with the Poiseuille pressure drop ($\Delta P$) and the "real" pressure drop ($\Delta P_{real}$) suggested by an earlier study for[12] symmetric airway bifurcations. The generations were numbered based on diameter [8] where length (L) and diameters (D) of each generation is available from a previous study[10]. Mass flow rate($\dot{m}$) and Reynolds number (Re) was calculated from the simulation. "C" is an empirical constant that equals 1.85 for symmetric bifurcations [12].

$$\Delta P = \frac{\dot{m}}{\rho} \frac{128 \mu L}{\pi D^4} \quad (1)$$

$$\Delta P_{real} = \frac{C}{4\sqrt{2}} \left( Re \frac{D}{L} \right)^{\frac{1}{2}} \Delta P \quad (2)$$

The pressure comparison results (Fig. 4) showed C values of 1.15 ± 0.13 (Mean ± SEM) would lead to agreement between our simulation results and Eq. 2 predictions. These values are 38% lower than the results when C=1.85 and are comparable with similar studies for asymmetric human airway models [4].

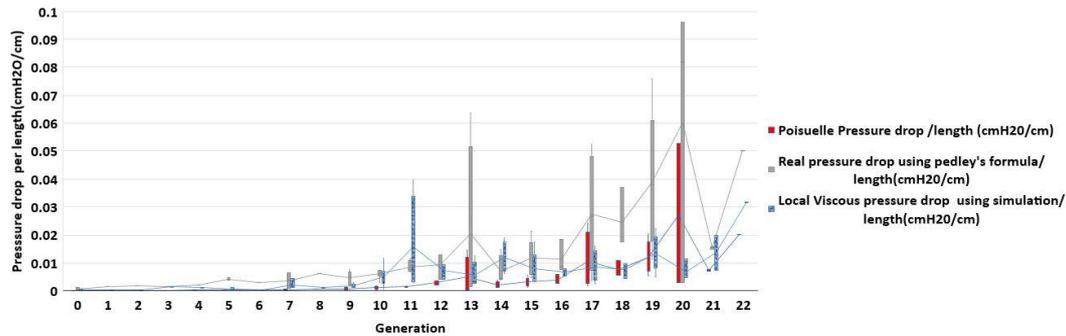

**Fig**. 4  Comparison of the pressure drop per unit length for inspiration (box and whiskers plot)[7]

## 4. CONCLUSIONS

The steady inspiratory and expiratory flow in a realistic a pig lung airway and symmetric bifurcations were numerically studied using CFD. The flow results in pig airway are discussed in relation to those of human airway models. The secondary flow results showed both similarities and differences compared to secondary flow patterns in simplified human airways[7]. Flow rate distribution during inspiration and expiration showed similar results, but secondary flows demonstrated significant differences that may translate to differences in flow-generated sounds. While the tracheal bronchus consumed 8% of the inlet flow, flow ratio between the right and left lungs were similar to those of human airways. The majority of the inlet flow remained in the major daughters due to the monopodial morphology of the pig lung airways. The inspiratory pressure drop results were in agreement with the predictions of Eqn. 2 when the empirical constant C was made comparable to those proposed for asymmetric human airways[4].

## ACKNOWLEDGMENT

This Study was supported by NIH R01 EB012142, R43HL099053.